\documentclass[prb,twocolumn,showpacs,preprintnumbers,amsmath,amssymb]{revtex4}
\usepackage{graphicx}
\usepackage{epsfig}

\def\olcite#1{[\onlinecite{#1}]}
\def\secref#1{section \ref{#1}}
\def\figref#1{figure \ref{#1}}

\def\Qminus{Q - }

\def\notq{{\overline{Q}}}
\def\notzero{{\overline{0}}}

\def\trp{\rm trp}
\def\exg{\rm exg}

\def\bcs{\Psi_{\rm BCS}}

\begin{document}

\title{Self-consistent T-matrix theory of superconductivity}

\author{B.~{\v S}op{\'\i}k$^{1,2}$, P.~Lipavsk\'y$^{1,2}$, M.~M{\"a}nnel$^{3}$, K.~Morawetz$^{3,4,5}$ and P.~Matlock$^6$}
\affiliation{$^1$Faculty of Mathematics and Physics, Charles University, Ke Karlovu 3, 12116 Prague 2, Czech Republic}
\affiliation{$^2$Institute of Physics, Academy of Sciences, Cukrovarnick\'a 10, 16253 Prague 6, Czech Republic}
\affiliation{$^3$M{\"u}nster University of Applied Science, Stegerwaldstrasse 39, 48565 Steinfurt, Germany}
\affiliation{$^4$International Institute of Physics (IIP), Federal University of Rio Grande do Norte, Av.~Odilon Gomes de Lima 1722, 59078-400 Natal, Brazil}
\affiliation{$^5$Max Planck Institute for the Physics of Complex Systems, 01187 Dresden, Germany}
\affiliation{$^6$Research Department, Universal Analytics Inc., Airdrie, AB, Canada}

\begin{abstract}
Using principles of the Fadeev-Lovelace-Watson multiple scattering
expansion, a \mbox{T-matrix} approximation is derived which coincides with
the Galitskii-Feynman T-matrix in the normal state, and yields the gap
in the superconducting state. Unlike other T-matrix approaches, the
theory satisfies not only the self-consistent Thouless criterion but
also the Baym-Kadanoff conditions for a conserving theory in
equilibrium. In single-mode approximation it simplifies to the
Eliashberg theory.
\end{abstract}

\pacs{71.10.-w, 74.20.-z, 03.75.Ss, 05.30.Fk}
\maketitle

\section{Introduction}

The known family of superconducting materials contains such diverse
systems as conventional metals and metallic alloys\cite{VIK82},
high-$T_{\rm c}$ ceramics \cite{Plak95}, fullerenes
\cite{RRGMHHPKZM91}, organic superconductors \cite{YXL04}, doped
diamond \cite{TNSTHK04}, heavy fermion metals \cite{Stewart84}, He-III
\cite{Lee97}, symmetric nuclear matter \cite{BS61,LSZ99,Bo03} and very
asymmetric nuclear matter in neutron stars \cite{YP04}, Fermi gases
\cite{BDZ08}, as well as hypothetical condensates like the colour
superconductivity of quarks \cite{ASRS08}. It is rather surprising how
many features of these systems have been successfully explained within
the mean-field Bardeen-Cooper-Schrieffer (BCS) theory and its
Green-function extension due to Eliashberg.  On the other hand, there
is a growing list of experimental facts which require the employment 
of more elaborate theories.

Recent theoretical approaches to superconductivity range from the trial
wave functions of Gutzwiller type \cite{EMGA06}, over improved
Eliashberg theories \cite{GPS95}, renormalization group approaches
within path integrals \cite{Shankar94}, exact diagonalizations and
quantum Monte Carlo studies \cite{DMOPR92} on simple models having
small size or infinite dimensions \cite{FJS93}, to the many-body
T-matrix approximations \cite{KM61,Patton71,MBL99,BGM01,HCCL07,M10,M11} and hybrid
theories combining the anomalous functions of Eliashberg type either
with the many-body T-matrix \cite{HRCZ07,HPZ09} or with the
fluctuation-exchange approximation \cite{PB95,Yan05,DS09}.

The formulation we shall present in the following is not of any of the
types mentioned above. We shall recall Watson's multiple scattering
theory;\cite{Watson53,GW64} his ideas were used by
Fadeev\cite{Fadeev60,Fadeev61,Fadeev62} and Lovelace\cite{Lovelace64}
in their description of few-body systems. In these small systems it is
crucial that each subsequent collision of a particle be with a
different partner; this is because the two-particle T-matrix covers
the binary interaction to all orders.\cite{JQ74}

This obvious physical principle is generally difficult to implement in
diagrammatic expansion methods, since generically the Feynman rules do
not impose any conditions on subsequent events; a summation obtains
over all possible partners.  Similarly, one does not find any
corresponding restriction of partners in the renormalization group
approaches.

A non-physical repetition of collisions with the same partner does not
introduce problems for normal metals, because the weight of repeated
collisions in unrestricted summations scales with the reciprocal
number of particles. This is in contrast to the case of
superconductors, where the condensate breaks this scaling behaviour
for pairing interactions and the non-physical repetition becomes a
serious problem. In \olcite{Lipavsky08} one of the present authors was
able to eliminate the repeated collisions from the Galtskii-Feynman
approximation\cite{Galitskii58} using Soven's concept of the effective
medium\cite{Soven67}. The Soven-type corrections are negligible in the
normal metal but become significant when the condensate develops.  The
approach proposed in \olcite{Lipavsky08} applies only to systems with
a non-retarded interaction. For many of the systems listed above,
however, the retarded nature of interaction is an essential ingredient
of a theoretical model if the goal is to achieve quantitative
agreement with experiment. The main focus of the present paper is to
derive a T-matrix approximation engineered for many-fermion systems
with pairing mediated by bosons; that is, with a retarded interaction
of finite range.

It is desirable that the theory be conserving in the Baym-Kadanoff
sense\cite{BK61}.  Methods which depend upon the introduction of
anomalous functions face great difficulty in this respect, as the
Baym-Kadanoff symmetry conditions are very restrictive. It may be
pointed out that anomalous functions themselves generically violate
particle-number conservation on the microscopic scale. It will be seen
shortly that in our formulation anomalous functions are not introduced
but instead appear as a consequence of other less disruptive
ingredients. Anomalous functions may be considered an
approximation of the two-particle Green function when the T-matrix
develops a singular separable term below the critical
temperature. This separable form constitutes a significant
simplification, confirming the vital r\^ole of anomalous functions in
the theory of superconductivity.

Theories starting with anomalous functions treat processes forming the
condensate nonperturbatively\cite{Bogolubov60,Haag62}, while other
processes are covered by low-order approximations.  In a construction
free of anomalous functions, all binary interactions may be described
to the same approximation, enabling the expression of exact
conservation laws.

It is known that the superconducting gap cannot be obtained in the
framework of what is referred to as a \emph{self-consistent}
Feynman diagrammatic expansion. It is also true that self-consistency
is a requirement of conserving theories. The problem
this poses is parallel to the conserving--gapless dichotomy in the
theory of Bose condensates\cite{HRCZ07}. In the approach of the
present paper, this problem is absent; the Baym-Kadanoff conditions
for conservation are satisfied, but not at the expense of the
superconducting gap.

In fact, Lorentz already in 1869 had some ideas which will guide us on
the correct path; the problem with the theory may be identified as the
presence of non-physical self-interactions, and the idea is to excise
these in a consistent way.  Looking at the issue from a different
viewpoint, the issue can be understood in terms of unphysical repeated
collisions; elimination of these is the major achievement of the
Fadeev-Lavelace-Watson multiple scattering approach, which is also
capable of producing a superconducting gap.\cite{Lipavsky08}
The first approach is more intuitive; the second approach supports
more rigorous justification. In this paper we present both before
showing that they are in fact equivalent.
We shall refer to the resulting theory as \emph{restricted self-consistent} or RSC.

The paper is organized as follows.
In \secref{DTA} we set the stage by reviewing the T-matrix approaches,
comparing the Galitskii-Feynman\cite{Galitskii58,BSI74} (GF) and the
Kadanoff-Martin\cite{KM61} (KM) approximations. We discuss the issue
of self-interactions in the GF approach, and the problem of the
Thouless criterion in the KM approach, thus illustrating the need for
a novel treatment suffering the problems of neither.
In \secref{RTMA} we introduce the idea of restricted self-consistency,
by which we intuitively construct a system of equations describing our 
new approach while avoiding the complexity of the multiple scattering theory;
the result is the RSC theory.
In \secref{AP} we begin to analyze this 
theory, showing the appearance of the gap and also the normal-state
coincidence with the GF theory.  The separable approximation of the
singular part of the T-matrix is shown to lead to the Eliashberg
theory.
In \secref{TPS} we prove that the two-particle Green function in the
RSC theory satisfies the conditions of Baym and Kadanoff 
for theories to be conserving on the microscopic level. 
Next, in \secref{MSA} we turn our attention to a derivation of
a T-matrix approximation from the multiple scattering theory.  After
this is established, it is shown that this actually amounts to a more
rigorous derivation of the same RSC theory constructed in
\secref{RTMA}.

\section{Diagrammatic T-matrix approaches and problems}
\label{DTA}

In this section we begin by reviewing some ideas about
self-consistency and self-interaction, and discuss the need for
restricted self-consistency.  We proceed to consider the concept of
self-interactions mediated by the condensate, the issue of repeated
collisions, and the problem of mutual exclusivity of self-consistency
and appearance of the gap. 
The T-matrix theory can either be constructed via a so-called partly
self-consistent or fully self-consistent diagrammatic expansion.  We
conclude this section by discussing both of these, and pointing out
why neither approach in fact produces a satisfactory theory -- a
problem which we will resolve in \secref{RTMA}.

\subsection{Lorentz self-consistency}

A standard starting point is the search for functionals of the bare or
dressed Green functions, $\Sigma[G_0]$ or $\Sigma[G]$. Actually,
neither $G_0$ nor $G$ is suited to describe a single collision
isolated from the series of collisions a given particle undergoes in
the many-body system. This problem of self-consistency was first
discussed by Lorentz in 1869. As his analysis is based on the
well-understood electric field and allows for a transparent
explanation, we review it here before applying such ideas to
superconductivity. Lorentz theory is covered in detail by chapter 13
of Kittel's textbook\cite{Kit71}.

Lorentz considered a gas of $N$ particles. The applied electric field
${\bf E}_0$ polarizes this gas so that the electric field ${\bf E}$
inside has a mean value given by 
${\bf E}_0=(1+\chi)\langle{\bf E}\rangle$, 
where $\langle{\bf E}\rangle$ denotes the field averaged over particle
configurations. We consider a conceptual correspondence; the applied
field corresponds to the bare line, the mean internal field to the
dressed line.

The internal field at point ${\bf r}$ is a sum of the applied field
and polarization fields of individual particles, 
${\bf E}({\bf r})= {\bf E}_0({\bf r})+\sum_{i=1}^N {\bf E}_i({\bf r})$. 
The polarization field of particle $i$ is
${\bf E}_i({\bf r})= M_i({\bf r}-{\bf r}_i){\bf E}^{(i)}({\bf r}_i)$,
determined by the field ${\bf E}^{(i)}({\bf r}_i)$ acting on this particle
and the tensor $M_i$ describing its polarizability and propagation of
the field. 

It is customary to assume that the field acting on a particle equals 
to the internal field, ${\bf E}^{(i)}({\bf r}_i)\approx{\bf E}
({\bf r}_i)$. Such step corresponds to the fully self-consistent
approximation; the 
internal field is taken as the only physically relevant quantity 
in the system. However, the polarization field diverges in the dipole approximation, 
${\bf E}^{(i)}({\bf r}_i)\to\infty$.
In the very dilute case one can remove the divergence using the
applied field, ${\bf E}^{(i)}({\bf r}_i)\approx{\bf E}_0({\bf r}_i)$,
and eventually add contributions of two, three and more particles.
This corresponds to the non-selfconsistent expansion. 

The mean field is also free of divergences, therefore it is plausible
to write ${\bf E}^{(i)}({\bf r}_i)\approx\langle{\bf E}({\bf  r}_i)\rangle$ 
as a basis of a convergent fully self-consistent approximation.  
Such an approximation amounts to the use of an averaged field as a
source in internal processes, and cannot generally be justified.
The correct prodecure would be to evaluate the electric field 
for each configuration and to perform the average only as a final step.

The solution proposed by Lorentz is simple and elegant. Since
the particle does not act on itself, it is correct 
to exclude its contribution, writing
${\bf E}^{(i)}({\bf r}_i)={\bf E}_0({\bf r}_i)+\sum_{j\neq i}^N {\bf E}_j({\bf r}_i)$. 
To close the set of equations one needs ${\bf E}^{(i)}$ as a function
of the mean internal field. To this end surrounding particles are represented
by the effective medium located everywhere except for the vicinity of 
the particle $i$. The field acting on the particle $i$ is then the field inside 
a spherical cavity, ${\bf E}^{(i)}({\bf r}_i)=\langle{\bf E}({\bf r}_i)\rangle
(1+\chi)/(1+{\frac23}\chi)$. 
In this way Lorentz achieves self-consistency, avoiding the action of
a given particle on itself.

One can adapt this Lorentz principle of self-consistency to the
interacting Fermi liquid in two different ways. First, one can view
fermions as Lorentz particles and their interaction as the internal
field. The Lorentz self-consistency then eliminates self-interaction;
this is discussed in \secref{RTMA}.  The second approach is slightly
more involved. The wave function of a selected particle plays the
r\^ole of the electric field propagating in the medium, and is
scattered by all other particles.  This approach, eliminating
non-physical repetition of collisions, will be discussed in
\secref{MSA}.

\subsection{Self-interaction mediated by the condensate}

\begin{figure}
\centerline{\psfig{file=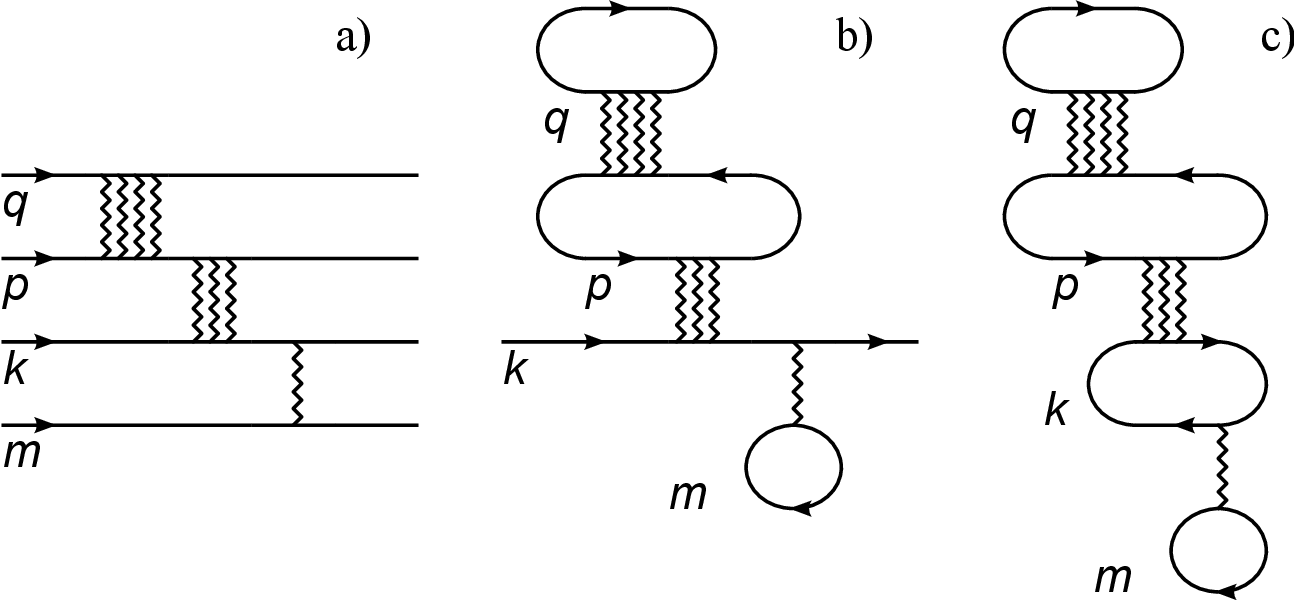,width=8cm}}
\caption{
Condensate-mediated self-interaction: 
Arrows represent bare fermionic Green functions $G_0$ and wavy lines are 
boson-mediated interactions. 
{\bf a)} Schematic picture of four interacting particles of
initial momenta $k$, $p$, $q$ and $m$. Due to the Pauli 
principle all particles are in different states, in particular $m\neq k$ 
and $q\neq k$, and $p\neq m$. {\bf b)} and {\bf c)} Corresponding Feynman 
diagrams for the Green function of the particle $k$ and for the 
thermodynamical potential. For lines closed into 
loops momenta are summed over with no restriction.}
\label{FigFourInteract}
\end{figure}

A self-interaction can be of various types; here we focus on a
self-interaction which is mediated by the condensate.  Before we
discuss this complex process, it is worthwhile to recall the simple
self-interaction appearing in the familiar context of the mean
field, when the true interaction is approximated by the scalar 
potential due to all electrons. In figures \ref{FigFourInteract}b
and~\ref{FigFourInteract}c one can see the lowest order of the 
mean-field potential given by the potential line and loop of 
summation index $m$. 

The relative error due to the mean-field self-interaction depends 
on the size of the system. In a single atom, each electron is 
bounded by a potential which asymptotically approaches the Coulomb 
potential of the remaining ion. In the mean-field approximation, 
however, the atom is neutral and the binding potential asymptotically 
approaches zero. In the infinite system with delocalized electrons the 
self-interaction is negligible. It becomes essential, however, when 
the infinite system contains bound states.

The mean-field self-interaction cancels with a corresponding
`self-exchange' of the Fock term. For details see
Appendix~\ref{Hartree}.  Although it is understood that eventually due
to higher-order diagrams all self-interactions will compensate each
other and the correct theory will emerge, such a formulation is not
viable for practical approximations. We will see in the next section
that it is far more profitable to reconsider the summation rule
itself, and to exclude the self-interaction directly as it is done in
the original theory of Hartree.

A self-interaction mediated by the condensate is shown in \figref{FigFourInteract}. 
The summation procedure does not respect that $m\ne k$, yielding the
Hartree self-interaction at $m=k$.  
Now let us focus on the two-loop contribution in the upper part, which
contains a more subtle self-interaction. We assume that $p\ne k$ which
is always guaranteed for separable potentials of BCS type coupling
only spin $\uparrow$ with spin $\downarrow$. Aside from their mutual
interaction, particles $p$ and $k$ also interact with the other
particles in the system. One such background particle is $q$, and the
sum over $q$ does not respect the condition $q\ne k$, leading to a
self-interation mediated by particle $p$. In the normal metal such
mediated processes are negligible, but in the superconductor the
condensate leads to enhancement of binary interactions with $p=-k$ and
$q=-p$. The weight of contributions with $q=k$ is thus finite,
so that such mediated processes may no longer be ignored.

We note that the $q$-loop merely dresses the $p$-line. Expansions
based on dressed Green functions do not include the diagram in
\figref{FigFourInteract}b; its contribution is hidden inside the
self-consistent scheme. In the non-selfconsistent expansion one can
eventually avoid this problem by demanding $q\ne k$. This implies that
the Green function in the $p$-loop is not dressed by all processes;
its value does not include interaction with the state $k$. It thus
becomes manifest that in the self-consistent expansion we need some
concept of restricted self-consistency.

\subsection{Repeated collisions}
Figure~\ref{FigFourInteract} includes two sequential interactions of
particle $k$ with particles $p$ and $m$. Since the T-matrix describes
the binary collision between $k$ and $p$ to infinite order, in the
subsequent collision the particle $k$ must actually encounter a
different, new partner. However, expanding the dressed Green function
in powers of the self-energy $G=G_0+G_0\Sigma G_0+G_0\Sigma G_0\Sigma
G_0+\ldots$, one can see that sequential events are described by
successive products of the self-energy $\Sigma$.  By definition, the
self-energy includes all processes and there is no restriction with
respect to the previous $\Sigma$ factor.

Mediated self-interactions and repeated collisions are 
closely connected concepts. Figure \ref{FigFourInteract}c shows
the diagram for the thermodynamic potential from which one can
generate the diagram of figure \ref{FigFourInteract}b. 
The contribution with $p=m$ can be classified either as a
self-interaction mediated by particle $k$, or as a repeated
collision along the propagation line of particle $k$.
Similarly, $q=k$ is either a mediated self-interaction, 
or a repeated collision on the $p$ line. 
We shall indeed find that a restriction of mediated
 self-interaction is in fact equivalent to an
elimination of repeated non-physical collisions.

\subsection{Dichotomy of self-consistency and gap}
\label{CGD}

In self-consistent approximations the self-energy $\Sigma[G]$ is a
functional of the dressed Green function $G$. The T-matrix
$T\sim\delta\Sigma/\delta G$ becomes divergent below the critical
temperature in the pairing channel; this is the element connecting a
particle of energy, momentum and spin $k=(\omega,{\bf k},\uparrow)$ with
its conjugate $-k=(-\omega,-{\bf k},\downarrow)$. Keeping the divergent
term only, $\Sigma(k)\approx T^{\rm div}G(-k)$, the Dyson equation
$G=G_0+G_0\Sigma G$ is easily solved giving 
$G=\left(1-\sqrt{1-4T^{\rm div}/(\omega^2-\xi_k^2)}\right)
(\omega+\xi_k)/(2T^{\rm div})$. This peculiar dressed Green function
exhibits no gap. In the bare Green function $G^{-1}_0=\omega-\xi_k$ we have
used a symmetric band structure $\xi_{-k}=\xi_k=k^2/2m-\mu$ for 
simplicity.

The gap easily emerges in the non-selfconsistent approximation
$\Sigma[G_0]$ with the T-matrix $T_0\sim\delta\Sigma/\delta G_0$.
Keeping the divergent term, $\Sigma(k)\approx T^{\rm div}_0 G_0(-k)$,
the Dyson equation results in the Gor'kov Green function
$G^{-1}=\omega-\xi_k-T_0^{\rm div}/(-\omega-\xi_k)$ with two
poles at $\omega=\pm\sqrt{\xi_k^2-T_0^{\rm div}}$.  The divergent
element of the T-matrix is a separable function which splits into
products of two gap functions $T_0^{\rm div} = - \Delta^*\Delta$.

The fundamental problems of the self-consistent approximation stem
from the scale dependence
of the Brillouin-Wigner self-consistent expansion scheme,
while the non-selfconsistent perturbative expansion of
Rayleigh-Schr{\"o}dinger type is size-consistent.\cite{Fulde} As
already mentioned, approximations which produce errors for few-body
systems do so also in the superconducting state because of
condensate-assisted processes.

The gap is a necessary part of any theory of superconductivity.  The
self-consistency is required only in some situations, for example the
Thouless criterion of the superconducting transition, or by the
closely related existence of a Goldstone mode.\cite{HRCZ07} What is
less well known is that the missing self-consistency also causes
trouble in microscopic studies of non-equilibrium superconductivity
beyond linear response. At some stage any study reaches the problem of
a `non-selfconsistent' distribution which is most commonly
circumvented by an (often implicit) assumption of local
equilibrium.

\subsection{Galitskii-Feynman versus Kadanoff-Martin}
\label{GFvKM}

\begin{figure}[h]
\centerline{\psfig{file=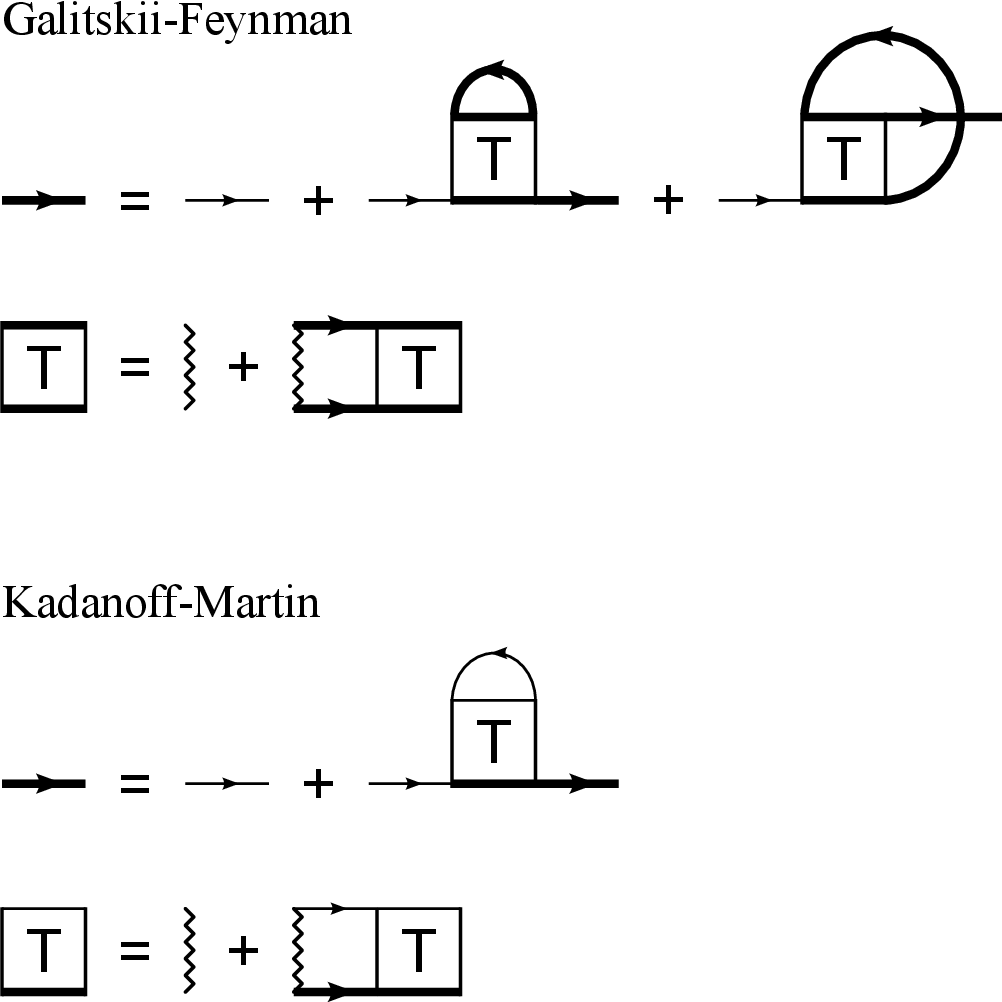,width=7cm}}
\caption{T-matrix approximations in diagrams: Both approximations have
  a self-energy constructed from the many-body T-matrix. The
  interaction carried by boson propagators, shown by wavy lines, is
  included in the ladder approximation.  Thick arrows are self-energy
  dressed Green functions, while thin arrows are bare Green functions.
}
\label{TmatrixGFKM}
\end{figure}

The GF and the KM approximations are compared in diagrammatic
representation in \figref{TmatrixGFKM}. Both are based on the
many-body T-matrix in the ladder approximation.  As one can see, the
KM approximation is nothing more than a simplified version of the GF
approximation, neglecting an exchange and possessing only a bare line
in the closed loop of self-energy, and in fact the exchange channel
contributes only if the particles have parallel spins. Nevertheless,
these two approximations are quite different with disjunct fields of
application.

The GF approximation is used in Nuclear Physics for both equilibrium
\cite{KM93,Bo99,Bo02} and non-equilibrium \cite{bm90,LSM97} problems,
in the theory of moderately dense gases \cite{B69} and liquid
\mbox{He-III} \cite{GBS76}, and in studies of electron-electron
correlations in molecules and solids
\cite{NIO06,Toulouse70,GKSB99,MPSSBR95,YSS07}.  There exist several
reasons why the KM approximation was never adapted to these problems.
The most important one is that the conservation laws are guaranteed
only if the T-matrix is symmetric with respect to the interchange of
the upper and the lower line of the intermediate propagators
\cite{BK61}; thus the lack of symmetry in the KM approximation is
viewed as unjustified and unacceptable.

The KM approximation is used exclusively in the theory of 
superconductivity\cite{MBL99,HCCL07,JML97,MJL99,CL08,CKL00,HCCL07b,CCHL07,HCCL08,CHCL08}. 
It describes the superconducting gap on the level of mean-field theory
and covers the lowest-order fluctuations. The GF approximation
cannot be employed for superconductors in spite of its superiority 
in other fields. Although it becomes unstable at the critical 
temperature \cite{BSI75} and the T-matrix diverges there, the GF
self-energy constructed from the T-matrix fails to describe the 
superconducting gap \cite{Wild60,KM61}. This is the general problem 
of self-consistent theories discussed in the previous subsection.

The paradox that the worse approximation (KM) yields the gap while the
better one (GF) fails in this regard was first noticed by
Prange\footnote{To our knowledge, Prange's results are not in print. 
Wild\cite{Wild60} and Tolmachev\cite{Tolmachev73} refer to them,
however.} and confirmed by Wild\cite{Wild60}, prior to
the work of Kadanoff and Martin \cite{KM61}. The Prange
paradox\cite{Tolmachev73} is not common knowledge and some authors,
see e.g.~\olcite{SB01}, report a superconducting gap obtained within
the GF approximation. Upon closer inspection one finds that in
simplification of some formulae, the bare Green function is used to close
loops,\cite{BGM01} a step which in fact turns the GF into the KM
approximation.

\subsection{Thouless criterion}
\label{Thcr}

The connection between formal perturbation theory and BCS-type
theories has been discussed by Thouless.\cite{Thouless60} It was found
that when the phase transition is approached from above, the critical
temperature can be determined through a criterion of stability of the
normal state; a divergence of the two-particle T-matrix signals the
transition. Thouless evaluated the T-matrix from non-selfconsistent
Green functions, but mentioned that corrections to the single-particle
functions are desirable.

The Thouless criterion also follows from Goldstone's
theorem.\cite{HRCZ07} The superconducting state is degenerate with
respect to the complex phase of the gap. According to Goldstone's
theorem there must be a corresponding branch of collective excitations
with energy going to zero in the long-wavelength limit.  The gap
appears as a divergence of the T-matrix at the frequency and momentum
of this Goldstone mode. For a constant complex phase, this divergence
is at zero frequency and momentum. At the critical temperature the
divergence gives the Thouless criterion for the nucleation of
superconductivity. The variational nature of this approach demands
that all Green functions in the T-matrix be
self-consistent.\cite{HRCZ07}

True self-consistency is essential. Beach, Gooding and Marsiglio
compared the self-consistent Thouless criterion with its
non-selfconsistent approximation in the attractive Hubbard
model.\cite{BGM01} They report that the non-selfconsistent criterion
yields a finite critical temperature while the self-consistent one
predicts a zero-temperature transition.  Following recent common use, 
by `Thouless criterion' we always mean its fully self-consistent form.

When formulated via the response to the complex phase modulation as
above, satisfaction of the Thouless criterion may be considered a
transport problem. Any system away from equilibrium requires
self-consistent distributions; the non-selfconsistent functions can be
used only under the assumption of local equilibrium, by which one
typically loses control over neglected terms. As an example we mention
the normal-current contribution to the time-dependent Ginzburg-Landau
equation, derived from the Thouless criterion\cite{LL08} and shown to
contribute already at the level of linear response\cite{LL09}.

It seems that neither the fully self-consistent nor the partly
non-selfconsistent diagrammatic approach can satisfy essential
theoretical criteria for a fundamental theory of superconductivity.
In the following section we introduce the RSC theory and begin to show 
how restricted self-consistency solves this problem.

\section{Eliminated self-interactions}

\label{RTMA}

We derive here a complete set of equations describing
superconductivity, which constitute the RSC theory. This is done via
an intuitive approach involving simple removal of the self-interaction
mediated by the condensate, thus allowing formation of the gap.

\subsection{Restricted self-energy}
\label{Redsel}
When two particles interact, 
their total energy and momentum $Q \equiv (\Omega,{\bf Q})$
is conserved; we may use this $Q$ to label binary processes. 
Dressing of a particle of four-momentum $k$ is given by the
self-energy $\Sigma_{\uparrow}(k)$ which is a sum over interacting
pairs,
\begin{equation}
\Sigma_{\uparrow}(k)=\sum_Q \left(\sigma_{Q\uparrow}(k)+
\sigma_{Q\uparrow}^{\trp}(k)+\sigma_{Q\uparrow}^{\exg}(k)
\right). 
\label{sigma}
\end{equation}
Here $\sigma_{Q\uparrow}(k)$ is a singlet contribution of single $Q$
and $\sum_Q\ldots\equiv\sum_{\Omega}\sum_{\bf Q}\ldots$ denotes sums
over bosonic Matsubara frequencies and discrete momenta in the
quantization volume $V$. Function $\sigma_{Q\uparrow}^{\trp}(k)$ is
a triplet contribution and $\sigma_{Q\uparrow}^{\exg}(k)$ is its
exchange counterpart. We assume singlet pairing and explicitly
treat only the singlet term.

The self-energy represents binary interactions averaged
over all possible many-body wave functions. This 
corresponds to the susceptibility in the Lorentz problem. Now we 
focus on the binary interaction in which the total four-momentum 
is $Q$. All other processes are treated as a background, represented
by a sum
\begin{equation}
\Sigma_{{\notq}\uparrow}(k)=\sum_{Q'\ne Q} \sigma_{Q'\uparrow}(k)+
\sum_{Q'}\left(\sigma_{Q'\uparrow}^{\trp}(k)+
\sigma_{Q'\uparrow}^{\exg}(k)\right)
\label{sigmared}
\end{equation}
over all modes but the $Q$-mode. The corresponding RSC Green function is
\begin{equation}
G_{{\notq}\uparrow}(k)=G^0_{\uparrow}(k)
+G^0_{\uparrow}(k)~\Sigma_{{\notq}\uparrow}(k)~ 
G_{{\notq}\uparrow}(k),
\label{Gq}
\end{equation}
where $G^0_{\uparrow}$ is the bare Green function.

In the spirit of the Lorentz approach we can also express the
restricted self-energy via a `cavity' in the effective medium 
\begin{equation}
\Sigma_{{\notq}\uparrow}(k)=\Sigma_{\uparrow}(k)-\sigma_{Q\uparrow}(k).
\label{sigmares}
\end{equation}
Since the dressed Green function is given by the Dyson equation 
\begin{equation}
\label{G}
G_{\uparrow}(k)=G^0_{\uparrow}(k)
+G^0_{\uparrow}(k)~\Sigma_{\uparrow}(k)~ 
G_{\uparrow}(k),
\end{equation}
we can express the RSC Green function via the dressed one 
\begin{equation}
\label{Gres}
G_{{\notq}\uparrow}(k)=G_{\uparrow}(k)-G_{\uparrow}(k)~
\sigma_{Q\uparrow}(k)~G_{{\notq}\uparrow}(k).
\end{equation}
This will allow us to close the self-consistency for the dressed
Green function avoiding the self-interaction and problems with the gap.

For the sake of clarity, we have written equations for only a selected 
spin orientation, the complementary equations are obtained simply by
flipping all spins.

\subsection{T-matrix}
\label{Tmatrix}
The contribution of the $Q$-mode to the self-energy reads
\begin{equation}
\sigma_{Q\uparrow}(k)=\frac{k_{\rm B}T}{V}{\mathcal{T}}_{\uparrow\downarrow}
(k,\Qminus k;k,\Qminus k)G_{{\notq}\downarrow}(\Qminus k)
\label{Sq}
\end{equation}
and similarly
\begin{equation}
\sigma_{Q\uparrow}^{\trp}(k)=\frac{k_{\rm B}T}{V}
{\mathcal{T}}_{\uparrow\uparrow}
(k,\Qminus k;k,\Qminus k)G_{\uparrow}(\Qminus k).
\label{Sqtr}
\end{equation}
The exchange channel reads
\begin{equation}
\sigma_{Q\uparrow}^{\exg}(k)=\frac{k_{\rm B}T}{V}
{\mathcal{T}}_{\uparrow\uparrow}
(k,\Qminus k;\Qminus k,k)G_{\uparrow}(\Qminus k).
\label{Sqex}
\end{equation}
We have used the RSC Green function 
$G_{{\notq}\downarrow}(\Qminus k)$ to close the loop of the singlet
channel \eqref{Sq}. In this way we have eliminated the interaction
of the $(\Qminus k;\downarrow)$-particle with the $(k;\uparrow)$-particle,
and therefore the mediated self-interaction of the $(k;\uparrow)$-particle.

To disallow self-interactions in intermediate processes 
the T-matrix must be constructed as
\begin{multline}
{\mathcal{T}}_{\uparrow\downarrow}(k,\Qminus k;p,\Qminus p)=
D(k,\Qminus k;p,\Qminus p)
\\
-\frac{k_{\rm B}T}{V}{\sum_{k'}}~
D(k,\Qminus k;k',\Qminus k')
\\
\times G_{\uparrow}(k')
G_{{\notq}\downarrow}(\Qminus k')
{\mathcal{T}}_{\uparrow\downarrow}(k',\Qminus k';p,\Qminus p),
\label{Tq}
\end{multline}
where $D$ is a bosonic interaction line with interaction vertices
included. The sum runs over momenta and fermionic Matsubara 
frequencies. We follow the sign convention of \olcite{AGD63} 
section 14.2. with $D$ becoming the interaction potential in the 
non-retarded limit.
The triplet T-matrix ${\mathcal{T}}_{\uparrow\uparrow}$
is analogous, with both Green functions dressed
\begin{multline}
{\mathcal{T}}_{\uparrow\uparrow}(k,\Qminus k;p,\Qminus p)=
D(k,\Qminus k;p,\Qminus p)
\\
-\frac{k_{\rm B}T}{V}{\sum_{k'}}~
D(k,\Qminus k;k',\Qminus k')
\\
\times G_{\uparrow}(k')
G_{\uparrow}(\Qminus k')
{\mathcal{T}}_{\uparrow\uparrow}(k',\Qminus k';p,\Qminus p).
\label{Tqtr}
\end{multline}

The set of equations is closed by the relation for the
density of particles
\begin{equation}
n_{\uparrow}=\frac{k_{\rm B}T}{V}\sum_k
G_{\uparrow}(k){\rm e}^{-i\omega\eta}
\label{nG}
\end{equation}
with $\eta$ infinitesimal and positive. This relation determines the
chemical potential; in equilibrium metals the electrons of spin
$\uparrow$ and $\downarrow$ have identical chemical potential, but in
transient systems two different chemical potentials might be defined
by this relation.

The set (\ref{sigma}-\ref{nG}) constitutes the RSC theory. 
This is the main result of the present
paper, and provides a complete description of superconductivity. We
will return to its derivation later in \secref{MSA}.
We take a moment to remind the reader that while this set of equations
describing the RSC theory can be represented diagrammatically, it must
be remembered that the usual Feynman rules have been modified.

\section{Gap and self-consistency}
\label{AP}

We now endeavour to show that the RSC theory constructed in the
preceding section not only yields the gap, but also satisfies the
Thouless criterion. We then prove that it exhibits the two-particle
symmetries which are the Baym and Kadanoff criteria for conserving
theories.

\subsection{Gap equation}
\label{Gapeq}

In the superconducting state there is a singlet channel in which the
T-matrix becomes singular. In equilibrium this divergence appears at
zero energy, $\Omega=0$, and in the absence of currents it is at zero
momentum, ${\bf Q}={\bf 0}$. This is the mode $Q=0\equiv (0,{\bf 0})$.
Its T-matrix is separable \cite{BSI74,BSI75} and diverges proportional
to the volume so that this single-mode contribution to the sum
is finite in the limits $V\to\infty$ and $T\to 0$;
\begin{equation}
\frac{k_{\rm B}T}{V}{\mathcal{T}}_{\uparrow\downarrow}(k,-k;p,-p)=-\phi^*(k)\phi(p)
\label{ap1}
.\end{equation}
The zero-mode contribution to the self-energy has a finite value
\begin{equation}
\sigma_{0\uparrow}(k)= - 
\phi^*(k)G_{{\notzero}\uparrow}(-k)\phi(k).
\label{ap2}
\end{equation}

According to \eqref{sigmares}, the self-energy is a
sum of the zero-mode contribution and the restricted self-energy,
\begin{equation}
\Sigma_{\uparrow}(k)=-
\phi^*(k)G_{{\notzero}\uparrow}(-k)\phi(k)+
\Sigma_{{\notzero}\uparrow}(k). 
\label{sigmadelta}
\end{equation}
The singularity thus does not enter the RSC Green function
$G_{{\notzero}\uparrow}$.  One can consider $G_{{\notzero}\uparrow}$
as the Green function of the normal metal.

Using \eqref{Gres}, the dressed Green function can be expressed via 
the RSC propagator
\begin{equation}
G_{\uparrow}(k)=G_{{\notzero}\uparrow}(k)-G_{{\notzero}\uparrow}(k)~
\phi^*(k)G_{{\notzero}\downarrow}(-k)\phi(k)~G_{\uparrow}(k).
\label{ap5}
\end{equation}
This equation shows that $\phi(k)$ equals the energy- and
momentum-dependent anomalous self-energy, which gives the
superconducting gap.

To connect with the Eliashberg theory we assume the system to be
symmetric in spins, $G^0_\downarrow(k)=G^0_\uparrow(k)$, and have no
supercurrent so that $G^0_\uparrow(\omega,-{\bf k})=
G^0_\uparrow(\omega,{\bf k})$. Splitting the restricted self-energy
into its even and odd parts
\begin{align}
\chi(k)&={\frac12}\big(\Sigma_{{\notzero}\uparrow}(k)+
\Sigma_{{\notzero}\downarrow}(-k)\big)
\label{el1}\\
\omega\big(Z(k)-1\big)&={\frac12}\big(\Sigma_{{\notzero}\uparrow}(k)-
\Sigma_{{\notzero}\downarrow}(-k)\big)
\label{el2}
\end{align}
one can express equation \eqref{ap5} as
\begin{eqnarray}
G_{\uparrow}^{-1}(k)&=&\omega -\xi_{\bf k}-\Sigma_{{\notzero}\uparrow}(k)+
\frac{\phi^*(k)\phi(k)}{-\omega -\xi_{\bf k}-\Sigma_{{\notzero}\downarrow}(-k)}
\nonumber\\
&=&\omega Z(k)-\xi_{\bf k}-\chi(k)
+\frac{\phi^*(k)\phi(k)}{ -\omega Z(k)-\xi_{\bf k}-\chi(k)}
\nonumber\\
\label{el3}
\end{eqnarray}
The gap in energy spectrum is sharp for real $\chi(k)$, when it has the 
renormalization familiar from the Eliashberg theory\cite{VIK82}
\begin{equation}
\Delta(k)={\phi(k)\over Z(k)}.
\label{el4}
\end{equation}

The anomalous self-energy itself follows from the equation for the
T-matrix \eqref{Tq} and the separability \eqref{ap1}
\begin{eqnarray}
\phi^*(k)&=&- \frac{k_{\rm B}T}{ V}
{\sum_{k'}}D(k,-k;k',-k')
\nonumber\\
&&~~~~~~~~~~\times G_{\uparrow}(k')G_{{\notzero}\downarrow}(-k')~\phi^*(k').
\label{ap6}
\end{eqnarray}
Deriving \eqref{ap6} we have used that $D/V\to 0$ in the thermodynamic
limit. This gap equation is a modified Eliashberg equation for the
off-diagonal self-energy.\footnote{Equation \eqref{ap6} corresponds
to the starting Green function equation of Elishberg. It is not 
averaged over the Fermi surface.}


At the critical line the gap vanishes and the nucleation kernel
approaches the normal state value, $DGG_{\notzero}\to DGG$. Here we
thus obtain the T-matrix made of fully self-consistent Green
functions.  We show that the RSC theory coincides with the GF theory
in the normal state of an infinite system. That it satisfies the
Thouless criterion discussed in Sec.~\ref{Thcr} is then a direct
consequence of this general limit.

In the normal state, the T-matrix has a finite value,
${\mathcal{T}}\sim D$.  According to \eqref{Sq} the single-mode
contribution to the self-energy vanishes in the thermodynamic limit,
$\sigma_{Q\uparrow}\propto 1/V\to 0$. The RSC Green function in this
case is equal to the dressed Green function, $G_{{\notq}\uparrow}\to G_{\uparrow}$, 
and the RSC theory may be identified with the GF approximation.

\subsection{Eliashberg equation}
\label{EE}
The Eliashberg equation is a simple approximation of the present RSC
theory: the reduced self-energy is approximated by the Migdal
self-energy,
\begin{align}
\Sigma_{{\notzero}\downarrow}(k)&\approx 
\Sigma^{\rm M}_{\downarrow}(k)
\nonumber\\
&=\frac{k_{\rm B}T}{V}\sum_QD(k,Q-k;Q-k,k)G_{\uparrow}(Q-k).
\label{Migdal}
\end{align} 
The Migdal self-energy is included in the T-matrix as its first-order
approximation, ${\mathcal{T}}\approx D$, of the exchange channel; one may
compare the summation in \eqref{Migdal} with expression \eqref{Sqex}.
It is easy to inlcude the singlet and direct-triplet channel at
first order since they yield the mean field of Hartree type. This
contribution is usually ignored for the phonon-mediated interaction,
however.

We see that in the superconducting state the RSC theory closely
parallels the Eliashberg theory, albeit with some differences. In the
RSC theory all processes, whether they be normal collisions or Cooper
pairing, are treated within the same T-matrix approximation. In the
Eliashberg theory the normal processes are in the Migdal approximation
while the pairing is covered by equations for the $\phi$ which is
described by the approximation corresponding to the T-matrix.

\section{Two-particle symmetry and conservation laws}
\label{TPS}
In this section we demonstrate that the RSC theory satisfies
symmetry conditions formulated by Baym and Kadanoff\cite{BK61} as
necessary for any theory to be conserving.  It is important to qualify
this by noting that these conditions alone are not sufficient; it
cannot thereby be claimed that the theory is conserving in the
Baym-Kadanoff sense, since the symmetries are actually required to
obtain \emph{in general}, while our RSC theory is limited to
\emph{equilibrium}.

\subsection{Baym-Kadanoff conditions}
\label{BKcon}
Let us rewrite both conditions of Baym and Kadanoff in the present
notation. The first BK condition states that the self-energy is linked to
the two-particle Green function $\mathcal{G}$ in two equivalent ways,
\begin{eqnarray}
&&\Sigma_{\uparrow}(k)G_{\uparrow}(k)=
\Big(\frac{k_{\rm B}T}{ V}\Big)^2\sum_{Q,p} D(k,\Qminus k;p,\Qminus p)
\nonumber\\
&&~~~~\times \big[{\mathcal{G}}_{\uparrow\uparrow}(p,\Qminus p;k,\Qminus k)+
{\mathcal{G}}_{\uparrow\downarrow}(p,\Qminus p;k,\Qminus k)\big],
\nonumber\\
\label{(A)1}\\
&&G_{\uparrow}(k)\Sigma_{\uparrow}(k)=
\Big(\frac{k_{\rm B}T}{ V}\Big)^2\sum_{Q,p} D(p,\Qminus p;k,\Qminus k)
\nonumber\\
&&~~~~\times \big[{\mathcal{G}}_{\uparrow\uparrow}(k,\Qminus k;p,\Qminus p)
+{\mathcal{G}}_{\uparrow\downarrow}(k,\Qminus k;p,\Qminus p)\big].
\nonumber\\
\label{(A)2}
\end{eqnarray}
The second BK condition demands that the two-particle Green function be
symmetric with respect to the interchange of the upper and lower
lines
\begin{align}
\mathcal{G}_{\uparrow\uparrow}(k,\Qminus k;p,\Qminus p) &=&
\mathcal{G}_{\uparrow\uparrow}(\Qminus k,k;\Qminus p,p),
\label{(B)}\\
\mathcal{G}_{\uparrow\downarrow}(k,\Qminus k;p,\Qminus p) &=&
\mathcal{G}_{\downarrow\uparrow}(\Qminus k,k;\Qminus p,p).
\label{ct5}
\end{align}

Though conditions (\ref{(A)1}-\ref{ct5}) provide in the equilibrium
case only limited indication of the validity of full conservation
laws, it is significant that the RSC theory passes this test; one can
easily show that the other theories we have mentioned fail to satisfy
the BK conditions even in equilibrium.  For example, the KM
approximation does not satisfy the Baym-Kadanoff criterion
\eqref{ct5}. It should be noted that the precursor of the present RSC
theory in \olcite{Lipavsky08} also fails to satisfy the symmetry in
\eqref{ct5}.

\subsection{Two-particle Green function}
The self-energy can be split into triplet and singlet channels, and
the symmetries for each contribution tested separately. The GF
approximation satisfies both Baym-Kadanoff conditions.\cite{BK61} The
triplet channel in our theory is the same as in the GF theory, and
therefore satisfies (\ref{(A)1}-\ref{(A)2}) and \eqref{(B)}. We thus
focus on the singlet channel in which the RSC theory differs from
the GF approximation.

The condition (\ref{(A)1}-\ref{(A)2}) links the single-particle Green 
function $G$ with the two-particle function $\mathcal{G}$. In the present
approximation the two-particle function is related to the T-matrix through
\begin{multline}
{\mathcal{G}}_{\uparrow\downarrow}(k,\Qminus k;p,\Qminus p)=
G_{\uparrow}(k)G_{{\notq}\downarrow}(\Qminus k)
\delta(k-p) \\
-G_{\uparrow}(k)G_{{\notq}\downarrow}(\Qminus k)
 G_{\uparrow}(p)G_{{\notq}\downarrow}(\Qminus p) \\
\times {\mathcal{T}}_{\uparrow\downarrow}(k,\Qminus k;p,\Qminus p)
\label{ct1}
\end{multline}
Substituting \eqref{ct1} into \eqref{(A)1} and \eqref{(A)2} one 
may check that both formulae yield the singlet self-energy given by
relations \eqref{sigma}, \eqref{Sq} and \eqref{Tq}. 

\subsection{Two-particle symmetry}

Condition \eqref{ct5} is somewhat nontrivial, demanding that the
singlet two-particle function be invariant under interchange of the
upper and lower lines. This symmetry is not obvious from expression
\eqref{ct1}.

First we show that the T-matrix \eqref{Tq} is symmetric with respect
to the interchange of the upper and lower lines
\begin{equation}
{\mathcal{T}}_{\uparrow\downarrow}(k,\Qminus k;p,\Qminus p)=
{\mathcal{T}}_{\downarrow\uparrow}(\Qminus k,k;\Qminus p,p),
\label{ct2}
\end{equation}
in spite of the fact that the upper line is constructed from RSC
Green functions while the lower line uses dressed Green functions. 

The T-matrix is a functional of the interaction ${\mathcal{T}}[D]$, which can
be expanded in powers. We prove the symmetry \eqref{ct2} to a general 
order $n$. First, we link powers of the T-matrix with powers of the 
two-particle Green function \eqref{ct1} using
\begin{equation}
{\mathcal{T}}_{\uparrow\downarrow}=D-\frac{V}{k_{\rm B}T}\sum 
D\cdot {\mathcal{G}}_{\uparrow\downarrow} \cdot D,
\label{TinD}
\end{equation}
which follows from \eqref{Tq} and \eqref{ct1}. 
The T-matrix to order $n$ in powers of $D$ thus depends on ${\mathcal{G}}$
to the power of $n-2$.

To prove the symmetry \eqref{ct2} we use induction. It is apparent
that the symmetry \eqref{ct2} is satisfied for two lowest orders
${\mathcal{T}}^{(1)}= D$ and 
${\mathcal{T}}^{(2)}= -D{\mathcal{G}}_{\uparrow\downarrow}^{(0)}D$, where 
${\mathcal{G}}_{\uparrow\downarrow}^{(0)}=G^0G^0$. We assume that the T-matrix
is symmetric up to order $n-2$ and show that the order $n-2$
two-particle Green function is then also symmetric. According to
relation \eqref{TinD}, this implies symmetry \eqref{ct2} to order $n$.

The $Q$-mode contribution to the self-energy \eqref{Sq} can be
rearranged as
\begin{eqnarray}
&&\sigma_{Q\uparrow}(k)G_{{\notq}\uparrow}(k) 
\nonumber\\
&&\qquad=\frac{k_{\rm B}T}{ V}{\mathcal{T}}_{\uparrow\downarrow}
(k,\Qminus k;k,\Qminus k)G_{{\notq}\downarrow}(\Qminus k)
G_{{\notq}\uparrow}(k)
\nonumber\\
&&\qquad=
\frac{k_{\rm B}T}{ V}{\mathcal{T}}_{\downarrow\uparrow}
(\Qminus k,k;\Qminus k,k)G_{{\notq}\downarrow}(\Qminus k)
G_{{\notq}\uparrow}(k)
\nonumber\\
&&\qquad=
G_{{\notq}\downarrow}(\Qminus k)
\sigma_{Q\downarrow}(\Qminus k).
\label{ct4}
\end{eqnarray}
This relation is based on symmetry \eqref{ct2}, therefore it is justified
to order $n-2$. 

Using relation \eqref{ct4} we can rearrange the product of two 
single-particle Green functions 
\begin{eqnarray}
&&G_{{\notq}\uparrow}(k)G_{\downarrow}(\Qminus k)
\nonumber \\
&&\quad=
G_{\uparrow}(k)
\big[1-\sigma_{Q\uparrow}(k)
G_{{\notq}\uparrow}(k)\big]
\nonumber\\
&&\quad\qquad\times
\big[1+G_{\downarrow}(\Qminus k)
\sigma_{Q\downarrow}(\Qminus k)\big]G_{{\notq}\downarrow}(\Qminus k)
\nonumber\\
&&\quad=
G_{\uparrow}(k)
G_{{\notq}\downarrow}(\Qminus k)
\big[1-\sigma_{Q\downarrow}(\Qminus k)
G_{{\notq}\downarrow}(\Qminus k)\big]
\nonumber\\
&&\quad\qquad\times
\big[1+G_{\downarrow}(\Qminus k)
\sigma_{Q\downarrow}(\Qminus k)\big]
\nonumber\\
&&\quad=G_{\uparrow}(k)G_{{\notq}\downarrow}(\Qminus k).
\label{ct3p1}
\end{eqnarray}
In the first step we have used \eqref{Gres} for Green functions
$G_{\downarrow}(\Qminus k)$ and $G_{{\notq}\uparrow}(k)$.
In the second step we have substituted from equation \eqref{ct4}. 
The last rearrangement follows again from \eqref{Gres}.

Using the relation \eqref{ct3p1} in equation \eqref{ct1} one finds
that from the symmetry of the T-matrix $\mathcal{T}$ follows the symmetry
of the two-particle Green function $\mathcal{G}$. We have thus proved that
from the symmetry of $\mathcal{T}$ up to order $n-2$ follows the symmetry
of $\mathcal{G}$ to the same order. Finally, using $\mathcal{G}$ symmetric up to
order $n-2$, from equation \eqref{TinD} one finds that the T-matrix is
symmetric up to order $n$. We have thus proved the symmetry
\eqref{ct2} to all orders.

From the symmetry of the T-matrix follows the symmetry \eqref{ct5} of
the two-particle Green function. In equilibrium, the RSC theory
thus satisfies the conditions of Baym and Kadanoff.

\section{Multiple scattering approach}
\label{MSA}

In the above derivation we have removed the self-interaction using the
idea of Lorentz regarding the interaction potential. The theory can in
fact be justified quite systematically using the Fadeev-Lovelace-Watson
multiple scattering
expansion\cite{Watson53,GW64,Fadeev60,Fadeev61,Fadeev62,Lovelace64,JQ74}
in which the Lorentz idea is applied to the wave function of a
particle. 
One may note that while the multiple scattering theory approach in
\olcite{Lipavsky08} was limited to non-retarded interactions, the
following presentation is applicable to a general interaction
mediated by bosons.

\subsection{Coherent propagation}
\label{ZAC}
In the multiple scattering theory one assumes that it is possible to
identify collisions of a selected particle. In the system of many
identical particles this is obscured by the presence of exchange
processes. Fortunately, we can trace the single-particle history in
coherent propagation, which is essential for the formation of the gap.

Expanding the dressed Green function \eqref{G} in powers of the 
self-energy, $G_{\uparrow}(k)=G^{0}_{\uparrow}(k)+G^{0}_{\uparrow}(k)
\Sigma_{\uparrow}(k)G^{0}_{\uparrow}(k)+G^{0}_{\uparrow}(k)\Sigma_{\uparrow}(k)
G^{0}_{\uparrow}(k)\Sigma_{\uparrow}(k)G^{0}_{\uparrow}(k)+
\ldots$ one can see that between interactions an electron returns to its 
starting state $(k,\uparrow)$. The Dyson equation thus describes only 
coherent propagation.

In the Feynman expansion one can associate each self-energy
contribution $\sigma_{Q\uparrow}(k)$ with an encounter of a particle
in state $(k,\uparrow)$ with a particle in state 
$(\Qminus k,\downarrow)$.  In coherent propagation both particles 
return to their initial states, as can be seen in the arguments of the
T-matrix in \eqref{Sq}. Following Landau we will call such encounters
\emph{zero-angle collisions}.

The product $G^0_{\uparrow}\Sigma_\uparrow G^0_{\uparrow}\Sigma_\uparrow G^0_{\uparrow}$ represents two 
subsequent zero-angle collisions. In the GF approximation such a product
includes terms $G^0_{\uparrow}(k)\sigma_{Q\uparrow}(k) G^0_{\uparrow}(k)\sigma_{Q\uparrow}(k) 
G^0_{\uparrow}(k)$ in which the particle in the $(k;\uparrow)$-state encounters the 
particle in the $(\Qminus k;\downarrow)$-state. Since after the first
encounter both particles returned to their initial states, the second
self-energy thus describes an encounter of the same pair of particles.
Such repeated zero-angle collision is in fact incompatible with the
T-matrix because its ladder approximation already covers the binary
interaction to infinite order. Finite states of a completed collision 
given by the T-matrix cannot serve as initial states for the
same process again.

\subsection{Effective medium}
\label{RC}
The repeated zero-angle collision is a double-count equivalent to the
molecule polarized by its own radiation in the Lorentz problem and we
can remove it with similar theoretical tools. Application of the
Lorentz idea to fermions was put forward by Watson\cite{Watson53,GW64}
who formulated the perturbative expansion in terms of binary
T-matrices showing that repeated collision must be avoided. His
multiple-scattering approach was further developed in two
directions. Fadeev\cite{Fadeev60,Fadeev61,Fadeev62} and
Lovelace\cite{Lovelace64} have worked with detailed applications to
small systems.  Soven\cite{Soven67,VKE68,V69} has applied the
multiple-scattering approach to scattering of electrons on static
random potential in alloys. We adopt Soven's concept of self-energy.

In parallel with the susceptibility, the self-energy can be viewed as
an auxiliary complex potential which represents the mean effect of
true collisions. Thus, instead of adding progressively more diagrams
we look for a condition which determines the self-energy from a physical
rather than mathematical viewpoint.

We focus on the $(k;\uparrow)$-particle making a zero-angle 
collision in the $Q$-mode. Briefly, we want to evaluate the
self-energy contribution $\sigma_{Q\uparrow}$. This process is 
described in detail by the T-matrix ${\mathcal{T}}_{\uparrow\downarrow}
(k,\Qminus k;k,\Qminus k)$. All other processes form an environment
in which this one happens and are thus covered on the level of 
effective medium. In the spirit of Lorentz cavity we subtract the
contribution $\sigma_{Q\uparrow}$ from the self-energy. The effective
medium is thus described by the restricted self-energy 
$\Sigma_{{\notq}\uparrow}$ for the $\uparrow$ component and by the 
complete self-energy $\Sigma_{\downarrow}$ for the $\downarrow$ component.

The T-matrix describes an interation of two particles to infinite
order. Since we focus on the $(k;\uparrow)$-particle, we have to
average over the probability to find a collision partner in the
$(\Qminus k;\downarrow)$-state,
\begin{equation}
s_{Q\uparrow}(k)=\frac{k_{\rm B}T}{\Omega}{\mathcal{T}}_{\uparrow\downarrow}
(k,\Qminus k;k,\Qminus k)G_{\downarrow}(\Qminus k).
\label{tGFq}
\end{equation}
The collision in the $Q$-mode is not to be repeated, therefore 
\begin{equation}
G_{\uparrow}(k)=G_{{\notq}\uparrow}(k)+
G_{{\notq}\uparrow}(k)s_{Q\uparrow}(k)
G_{{\notq}\uparrow}(k).
\label{ct6}
\end{equation}
The scattering equation \eqref{ct6} defines the self-energy
indirectly. Comparing \eqref{ct6} with \eqref{Gres}
we find that the $Q$-mode contribution to the self-energy is given by
\begin{equation}
\frac{\sigma_{Q\uparrow}(k)}{ 1-\sigma_{Q\uparrow}(k)G_{{\notq}\uparrow}(k)}
=s_{Q\uparrow}(k).
\label{ct7}
\end{equation}

Finally we need the many-body T-matrix. Since our
background is described by $\Sigma_{{\notq}\uparrow}$ and
$\Sigma_{\downarrow}$, the ladder approximation of the many-body 
T-matrix 
\begin{multline}
{\mathcal{T}}_{\uparrow\downarrow}(k,\Qminus k;p,\Qminus p)=
D(k,\Qminus k;p,\Qminus p)
\\
-\frac{k_{\rm B}T}{\Omega}{\sum_{k'}}~
D(k,\Qminus k;k',\Qminus k')
\\
\times G_{{\notq}\uparrow}(k')
G_{\downarrow}(\Qminus k')
{\mathcal{T}}_{\uparrow\downarrow}(k',\Qminus k';p,\Qminus p)
\label{TqMS}
\end{multline}
is constructed from $G_{{\notq}\uparrow}$ and $G_{\downarrow}$.
The set of equations (\ref{sigma}-\ref{Gres}),  
(\ref{Sqtr}-\ref{Sqex}), \eqref{Tqtr} and (\ref{ct7}-\ref{TqMS}) is closed. 

\subsection{Relation to the eliminated self-interaction}
\label{RES}

The set of equations (\ref{sigma}-\ref{Gres}),  
(\ref{Sqtr}-\ref{Sqex}), \eqref{Tqtr} and (\ref{ct7}-\ref{TqMS}) is in fact
equivalent to the set of equations (\ref{sigma}-\ref{Tqtr})
which we derived intuitively earlier in the paper, and which define the RSC theory.
To see this, we use the symmetry \eqref{ct3p1} and readily rewrite
\eqref{TqMS} to obtain \eqref{Tq}. The two definitions of the T-matrix
are thus equivalent.

It remains to show that the self-energy is identical. From equation \eqref{ct7} we find
\begin{eqnarray}
\sigma_{Q\uparrow}(k)&=& s_{Q\uparrow}(k)\left(1-\sigma_{Q\uparrow}(k)G_{{\notq}\uparrow}(k)\right)
\nonumber\\
&=&s_{Q\uparrow}(k)\left(1-\sigma_{Q\downarrow}(\Qminus k)G_{{\notq}\downarrow}(\Qminus k)\right)
\nonumber\\
&=&\frac{k_{\rm B}T}{\Omega}{\mathcal{T}}_{\uparrow\downarrow}
(k,\Qminus k;k,\Qminus k)
\nonumber\\
&&\times G_{\downarrow}(\Qminus k)
\left(1-\sigma_{Q\downarrow}(\Qminus k)G_{{\notq}\downarrow}(\Qminus k)\right)
\nonumber\\
&=&\frac{k_{\rm B}T}{\Omega}{\mathcal{T}}_{\uparrow\downarrow}
(k,\Qminus k;k,\Qminus k)G_{{\notq}\downarrow}(\Qminus k), \nonumber\\
\label{tGFqsim}
\end{eqnarray}
therefore the expressions \eqref{ct7} and \eqref{Sq} yield the same
self-energy.  In the rearrangement we have used \eqref{ct4} and
\eqref{tGFq}.

\subsection{Comments on choice of restriction}

Finally we want to comment on the relation of the RSC theory to the
derivation in \olcite{Lipavsky08}. Here we identify the mode via
energy and momentum $Q\equiv(\omega,{\bf Q})$. In \olcite{Lipavsky08}
the mode was identified only via momentum ${\bf Q}$, which applies
only to non-retarded interactions and leads to different results.

In particular, the identification of a mode via momentum does not
provide a two-particle Green function symmetric with respect to
interchange of the upper and lower lines. The theory in
\olcite{Lipavsky08} thus does not satisfy the condition \eqref{ct5} of
Baym and Kadanoff and cannot be converted into a more convenient form
with restricted self-consistent Green functions in the closed loop.

Apparently, one can derive a theory with restricted self-consistency
in the loop and the mode identified via momentum ${\bf Q}$ by
elimination of mediated self-interactions in a manner similar to the
one employed in \secref{Redsel}. A set of equations obtained in this
way is not identical to the theory in \olcite{Lipavsky08}.
Differences are minor, however. The two approaches become identical in
the single-mode approximation leading to the same equation of BCS
type.

The RSC theory is restricted to equilibrium. In contrast, the theory
in \olcite{Lipavsky08} is based exclusively on double-time functions,
which allows one to extend it to non-equilibrium systems using either
Kadanoff-Baym or Keldysh machinery.

Extension of the present RSC theory cannot be achieved by a
straightforward application of the Kadanoff-Baym method. This is
because the $Q$-mode contribution $\sigma_{Q\uparrow}(k)$, depends on
two four-momenta, bosonic $Q\equiv(\Omega,{\bf Q})$ and fermionic
$k\equiv(\omega,k)$. Functions of two frequencies correspond to
three-time functions which have six analytic parts in the
non-equilibrium extension. This makes the putative non-equilibrium
version prohibitively complicated.

\section{Summary and conclusion}
\label{ConDis}
Self-consistent theories are unviable for superconductivity, as they
cannot yield a superconducting gap. So-called non-selfconsistent
approaches produce a gap, but can be shown to be non-conserving,
failing to satisfy the necessary Baym-Kadanoff conditions.  Applying
principles of the multiple-scattering theory to the T-matrix
approximation, we have derived a theory which describes the
superconducting gap, the structure of this theory being similar to a
renormalized Kadanoff-Martin approximation, but sporting two major
improvements. First, in the normal state the well-tested
Galitskii-Feynman approximation is recovered. Since the
Galitskii-Feynman T-matrix depends on self-consistent propagators, the
RSC theory satisfies the Thouless criterion. Second, the two-particle
propagator is symmetric with respect to interchange of the two lines
in its defining Feynman diagram. This symmetry allows the RSC theory
to satisfy the Baym-Kadanoff requirements for a conserving theory.
Finally, though the RSC theory may be approximated by the Eliashberg
theory, it may be noted that due to the more elaborate
self-consistency of the RSC theory, superconductivity conditions in
strongly-interacting systems are likely to be different from the
Eliashberg theory.

\begin{acknowledgments}
This work was supported by research plans MSM 0021620834 and
AV0Z10100521, grant projects GA\v{C}R 204/10/0687, 204/10/0212 
and 204/11/0015, DAAD-PPP (BMBF) Germany, PPP Taiwan, and by 
DGF-CNPq project 444BRA-113/57/0-1. The financial support by the 
Brazilian Ministry of Science and Technology is acknowledged.
\end{acknowledgments}

\appendix
\section{Mean-field self-interaction and Hartree approximation}
\label{Hartree}
This appendix follows the introductory part of Slater's 
paper\cite{Slater51} in which was simplified the Hartree-Fock 
method, constructing the basis of the Local-Density Approximation. 
Two simplifications are adopted within this section. First, we
assume a ground state of $N$ particles described by a single 
many-body wave function. Second, the interaction potential is 
of the Coulomb type. The Hamiltonian is thus a sum of the 
single-particle part and the interaction, 
$H=\sum_i H^{(1)}(x_i) + \sum_{k<i}V(x_i-x_k)$. 

\subsection{Hartree equations}
The Hartree equations are obtained by minimizing the energy
on the class of separable wave-functions of the form 
\begin{align}
W_{\rm H}=\int\!dx_1\!\ldots\! dx_2 \bar\psi_1(x_1)\!\ldots\! 
\bar\psi_N(x_N)H\psi_N(x_N)\!\ldots\!\psi_1(x_1),
\label{eH1}
\end{align}
where $x_i$ are coordinates, and a summation over spins is understood.
Varying the $\psi$ functions in the Hartree energy \eqref{eH1} one finds
\begin{align}
E_i\psi_i(x)&=
H^{(1)}\psi_i(x)
\nonumber\\
&+\left[\sum_{k\not=i}\int dx'
\bar\psi_k(x')\psi_k(x')V(x-x')\right]\psi_i(x).
\label{eH2}
\end{align}
Since the particle does not interact with itself, the term with $k=i$
is excluded from the sum. 

The mean potential
\begin{align}
\phi(x)=\sum_{k}\int dx'
\bar\psi_k(x')\psi_k(x')V(x-x')
\label{eH3}
\end{align}
includes contributions from all electrons. The Hartree equations
can be written in terms of the mean potential as
\begin{align}
E_i\psi_i(x)&=
H^{(1)}\psi_i(x)+\phi(x)\psi_i(x)
\nonumber\\
&-\left[\int dx'
\bar\psi_i(x')\psi_i(x')V(x-x')\right]\psi_i(x).
\label{eH4}
\end{align}
Briefly, the Hartree approximation is given by the mean potential
corrected by the self-interaction. 

\subsection{Hartree-Fock equations}
The Hartree-Fock equations are obtained by minimizing the energy,
on the class of anti-symmetrized separable functions of the form
\begin{align}
W_{\rm HF}=&{1\over N!}
\int dx_1\ldots dx_2
\nonumber\\
\times&\left|
\begin{array}{rcl}
\bar\psi_N(x_N)\!\!&\!\ldots\!\! &\!\bar\psi_N(x_1)\\
\ldots\ldots\!\! &\!\ldots\!\! &\!\ldots\ldots\\
\bar\psi_1(x_N)\!\!&\!\ldots\!\! &\!\bar\psi_1(x_1)
           \end{array}
\right|H
\left|\begin{array}{rcl}
\psi_1(x_1)\!\!&\!\ldots\!\! &\!\psi_1(x_N)\\
\ldots\ldots\!\! &\!\ldots\!\! &\!\ldots\ldots\\
\psi_N(x_1)\!\!&\!\ldots\!\! &\!\psi_N(x_N)
           \end{array}
\right|.
\label{eH5}
\end{align}
Unlike in Hartree's case, the $\psi$ functions in in Slater's determinant
may be assumed to be orthogonal without loss of generality.

Varying the $\psi$ functions in the energy \eqref{eH5} one finds 
\begin{align}
E_i\psi_i(x)&=
H^{(1)}\psi_i(x)+\phi(x)\psi_i(x)
\nonumber\\
&-\sum_{k}\left[\int dx'
\bar\psi_k(x')\psi_i(x')V(x-x')\right]\psi_k(x).
\label{eH6}
\end{align}
The last term is due to the exchange of particles and 
it is customary to refer to it as the Fock potential. In this 
spirit the mean potential $\phi$ is often called the 
Hartree potential. 

Note that Fock term includes a $k=i$ contribution, therefore
the self-interaction of the mean potential $\phi$ cancels
with the self-exchange. 

The Fock term corresponds to a single-electron charge. 
This can be seen from an effective density
\begin{align}
n_{i,x}(x')&=\sum_{k}
{\bar\psi_i(x)\bar\psi_k(x')\psi_k(x)\psi_i(x')\over
\bar\psi_i(x)\psi_i(x)}
\label{eH7}
\end{align}
in terms of which the Hartree-Fock equations are reminiscent of
usual single-particle Schr{\"o}dinger equation
\begin{align}
E_i\psi_i(x)&=
H^{(1)}\psi_i(x)+\phi(x)\psi_i(x)
\nonumber\\
&-\left[\int dx'n_{i,x}(x')V(x-x')\right]\psi_i(x).
\label{eH8}
\end{align}
The effective density corresponds to a single particle;
\begin{align}
\int dx'n_{i,x}(x')=1
\label{eH9}
,
\end{align}
as one finds integrating and summing the right hand side of
\eqref{eH7}. From orthogonality of the $\psi$ functions follows 
that only the term with $k=i$ contributes.

\section{Model of reduced interaction}
The BCS wave function
\begin{align}
\big|\bcs\big\rangle=\prod_{\bf k}\left(u_{\bf k}+v_{\bf k}
\psi^\dagger_{{\bf p}\uparrow}\psi^\dagger_{-{\bf p}\downarrow}\right)
\big|0\big\rangle
\label{ea1}
\end{align}
is known to be the exact ground state in the limit of infinite volume 
for the reduced interaction
\begin{align}
&\hat D=-\frac{\lambda}{ V}\sum_{{\bf k,p}}
\psi^\dagger_{{\bf p}\uparrow}
\psi^\dagger_{-{\bf p}\downarrow}
\zeta_{\bf p}\zeta_{\bf k}
\psi_{-{\bf k}\downarrow}
\psi_{{\bf k}\uparrow},
\label{ea2}
\end{align}
that is,
\begin{align}
D_{\uparrow\downarrow}(k,Q-k;p,Q-p)&=-\lambda\,\zeta_{\bf k}\zeta_{\bf p}
\delta_{{\bf Q},{\bf 0}}   
\nonumber\\
D_{\uparrow\uparrow}&=0.
\label{ea2x}
\end{align}
The $\zeta$ factors are form factors; $\zeta$ can be either a simple cutoff,
e.g. $\zeta_{\bf k}=\theta(\omega_{\rm c}-|\xi_{\bf k}|)$, or a more 
involved function covering nontrivial gap symmetries.
We test the present approximation against this exact result.

\subsection{Anomalous functions}
The BCS state implies mean-field approximation of the self-energy.
We will discuss all approximations in the time representation, in
which the Green function is the mean value of time-ordered product 
of field operators in different times,
\begin{align}
 G_\uparrow(t,{\bf k})&=-i\big\langle{\bcs}\big|{\sf T}\psi_{{\bf k}\uparrow}(t)
\psi_{{\bf k}\uparrow}^\dagger(0)\big|{\bcs}\big\rangle 
\nonumber\\
&\equiv -i\big\langle \psi_{{\bf k}\uparrow}\psi_{{\bf k}\uparrow}^{\dagger 0}\big\rangle.
\label{ea3}
\end{align}
From $G_0^{-1}=i\partial_t-\xi_{\bf k}$ follows
\begin{align}
 G_0^{-1}G_\uparrow&=\delta(t)+\big\langle[\hat D,\psi_{{\bf k}\uparrow}]
\psi_{{\bf k}\uparrow}^{\dagger0}\big\rangle 
\nonumber\\
&=\delta(t)+\zeta_{\bf k}\frac{\lambda}{ V}\sum_{\bf p}\zeta_{\bf p}
\big\langle\psi_{{\bf -k}\downarrow}^{\dagger}\psi_{{\bf -p}\downarrow}
\psi_{{\bf p}\uparrow}\psi_{{\bf k}\uparrow}^{\dagger0}\big\rangle.
\label{ea4}
\end{align}

The two-particle Green function exactly satisfies the anomalous 
decoupling 
\begin{align}
\big\langle\psi_{{\bf -k}\downarrow}^{\dagger}\psi_{{\bf -p}\downarrow}
\psi_{{\bf p}\uparrow}\psi_{{\bf k}\uparrow}^{\dagger0}\big\rangle
=& \big\langle\psi_{{\bf -k}\downarrow}^{\dagger}\psi_{{\bf -p}\downarrow}
\big\rangle\big\langle \psi_{{\bf p}\uparrow}\psi_{{\bf k}\uparrow}^{\dagger0}
\big\rangle
\nonumber\\
+&\big\langle\psi_{{\bf -k}\downarrow}^{\dagger}\psi_{{\bf k}\uparrow}^{\dagger0}
\big\rangle\big\langle \psi_{{\bf -p}\downarrow}\psi_{{\bf p}\uparrow}
\big\rangle .
\label{ea5}
\end{align}
This is easily proved using the Bogoliubov-Valutin transformation
$\psi_{{\bf k}\uparrow}=u_{\bf k}\gamma_{\bf k}+v_{\bf k}\beta_{\bf k}^\dagger $, 
and $\psi_{{\bf -k}\downarrow}
=u_{\bf k}\beta_{\bf k}-v_{\bf k}\gamma_{\bf k}^\dagger$, 
where $\beta$ and $\gamma$ are annihilation operators of excitations above 
the BCS state, $\beta_{\bf k}|{\bcs}\rangle=0$ and 
$\gamma_{\bf k}|{\bcs}\rangle=0$. Using the anticommutation relation
$\gamma_{\bf k}\gamma_{\bf p}^\dagger+
\gamma_{\bf p}^\dagger\gamma_{\bf k}=\delta_{\bf k,p}$ between operators
at equal times, one finds that both sides of \eqref{ea5} equal
$-v_{\bf k}u_{\bf k}v_{\bf p}u_{\bf p}\left(1-\delta_{\bf k,p}\right)
\big\langle\gamma_{\bf k}\gamma_{\bf k}^{\dagger0}\big\rangle$. 

By decoupling \eqref{ea5} one readily converts the non-perturbative 
equation of motion \eqref{ea4} into the mean-field equation of Gor'kov
type. The mean-field approximation for anomalous functions thus yields 
an exact solution for the reduced interaction \eqref{ea1}. 

Now we show that restricted self-consistency also yields the exact
solution. To this end we will compare our equation for the RSC
T-matrix with Gor'kov equations.

Let us first write down the self-energy following from the Gor'kov
theory. The product of normal mean values is proportional to $\delta_{\bf k,p}$. 
In the limit of infinite volume the contribution of this term to the 
interaction term in \eqref{ea4} vanishes as $1/V$ and only the product 
of anomalous functions survives
\begin{align}
G_0^{-1}G_\uparrow
&=\delta(t)+\zeta_{\bf k}\frac{\lambda}{ V}\sum_{\bf p}\zeta_{\bf p}
\big\langle \psi_{{\bf -p}\downarrow}\psi_{{\bf p}\uparrow}\big\rangle
\big\langle\psi_{{\bf -k}\downarrow}^{\dagger}\psi_{{\bf k}\uparrow}^{\dagger0}
\big\rangle .
\label{ea6}
\end{align}
We denote the anomalous Green function 
\begin{align}
F^*(t;{\bf k})=
\big\langle\psi_{{\bf -k}\downarrow}^{\dagger}\psi_{{\bf k}\uparrow}^{\dagger0}
\big\rangle 
\label{ea7}
\end{align}
and the gap function 
\begin{align}
\Delta_{\bf k}=
\zeta_{\bf k}\frac{\lambda}{ V}\sum_{\bf p}\zeta_{\bf p}
\big\langle \psi_{{\bf -p}\downarrow}\psi_{{\bf p}\uparrow}\big\rangle
\label{ea8}
\end{align}
in terms of which equation \eqref{ea6} reads 
\begin{align}
G_0^{-1}G_\uparrow
&=\delta(t)+\Delta_{\bf k}F^*.
\label{ea9}
\end{align}

Derivation of the equation for $F^*$ is similar to the above derivation
of equation \eqref{ea9}. It gives
\begin{align}
\tilde G_0^{-1}F^*
&=-\Delta_{\bf k}^*G_\uparrow,
\label{ea10}
\end{align}
where $\tilde G_0^{-1}=-i\partial_t-\xi_{\bf -k}$. 
The singular term $\delta(t)$ does not appear, as creation operators anticommute.
The $\Delta^*$ is obtained from 
\begin{align}
\Delta^*_{\bf k}=
\zeta_{\bf k}\frac{\lambda}{ V}\sum_{\bf p}\zeta_{\bf p}F^*_{\bf p}, 
\label{ea8hc}
\end{align}
which is the hermitian
conjugate of equation \eqref{ea8}. Substituting the solution of \eqref{ea10}
in \eqref{ea9} we find
\begin{align}
G_0^{-1}G_\uparrow
&=\delta(t)-\Delta_{\bf k}\tilde G^0_\downarrow\Delta_{\bf k}^* G_\uparrow.
\label{ea11}
\end{align}

The Fourier transformation of Eq.~\eqref{ea11} in time reads
\begin{align}
\left(\omega-\xi_{\bf k}\right)G_\uparrow
=1-\Delta_{\bf k}\left(-\omega-\xi_{\bf -k}\right)^{-1}\Delta_{\bf k}^* 
G_\uparrow.
\label{ea12}
\end{align}
The self-energy defined as $G_\uparrow^{-1}=\omega-\xi_{\bf k}-
\Sigma_\uparrow$, can be expressed in terms of the gap function as
\begin{align}
\Sigma_\uparrow(k)
&=-\Delta_{\bf k} G^0_\downarrow(-k)\Delta_{\bf k}^*.
\label{ea13}
\end{align}
where $ G^0_\downarrow(-k)=\left(-\omega-\xi_{\bf -k}\right)^{-1}$. This self-energy
yields the exact Green function for the infinite system with reduced 
interaction. 

Finally, we write down an explicit gap equation. Using the
anomalous Green function from \eqref{ea10} in the gap 
equation \eqref{ea8hc} one finds
\begin{align}
\Delta^*_{\bf k}=-\lambda\zeta_{\bf k}\frac{k_{\rm B}T}{ V}\sum_{{\bf p},\omega}\zeta_{\bf p}
G^0_\downarrow(-\omega,-{\bf p})\Delta^*_{\bf p}G_\uparrow(\omega,{\bf p}).
\label{ea21gap}
\end{align}
We have evaluated the equal-time Green function $F^*$ needed
in the gap equation by summing over Matsubara frequencies.

\subsection{Restricted self-consistent T-matrix}
We compare the self-energy \eqref{ea13} and the gap equation
\eqref{ea21gap} with their couterparts derived from the restricted
self-consistent T-matrix.

In treating the RSC T-matrix we will not benefit from anomalous
decoupling, but evaluate the resulting self-energy directly from the
above set of equations (\ref{sigma}-\ref{Tqtr}) with the reduced
interaction \eqref{ea2x}.

In equations \eqref{Tq} and \eqref{Tqtr} we must include the
spin-dependence of the interaction line. Substituting
$D_{\uparrow\uparrow}$ for $D$ in \eqref{Tqtr} we obtain that the
triplet T-matrix vanishes; ${\mathcal{T}}_{\uparrow\uparrow}=0$. The
triplet and exchange self-energies \eqref{Sqtr} and \eqref{Sqex} are
thus trivial, $\sigma^{\trp}_{Q\uparrow}(k)=0$ and
$\sigma^{\exg}_{Q\uparrow}(k)=0$.

With $D=D_{\uparrow\downarrow}$ the ladder equation \eqref{Tq} yields
a non-zero singlet T-matrix only for ${\bf Q}=0$. The self-energy is
thus a sum over only Matsubara frequencies;
\begin{align}
\Sigma_\uparrow(\omega,{\bf k})&=\frac{k_{\rm B}T}{ V}
\sum_\Omega 
{\mathcal{T}}_{\uparrow\downarrow}
\left(\omega,{\bf k},\Omega\!-\!\omega,\!-\!{\bf k};
\omega,{\bf k},\Omega\!-\!\omega,\!-\!{\bf k}\right)
\nonumber\\
&~~~~~~~~~~~~\times G_{\notq\downarrow}(\Omega\!-\!\omega,\!-\!{\bf k}).
\label{ea15}
\end{align}

The restricted self-energy 
\begin{align}
\Sigma_{\notzero\uparrow}(\omega,{\bf k})&=\frac{k_{\rm B}T}{ V}
\sum_{\Omega\not= 0} 
{\mathcal{T}}_{\uparrow\downarrow}
\left(\omega,{\bf k},\Omega\!-\!\omega,\!-\!{\bf k};
\omega,{\bf k},\Omega\!-\!\omega,\!-\!{\bf k}\right)
\nonumber\\
&~~~~~~~~~~~~\times G_{\notq\downarrow}(\Omega\!-\!\omega,\!-\!{\bf k})
\label{ea16}
\end{align}
has no singularity proportional to the volume and there is no sum over 
momenta, therefore it vanishes in the limit of infinite volume $V\to\infty$
\begin{align}
\Sigma_{\notzero\uparrow}(\omega,{\bf k})&=0.
\label{ea17}
\end{align}
The contribution of the zero Matsubara frequency, 
$Q=0\equiv (0,{\bf 0})$, is enhanced by singularity 
of Bose-Einstein statistics at condensates, therefore
\begin{align}
\Sigma_\uparrow(k)&=
\frac{k_{\rm B}T}{V}
{\mathcal{T}}_{\uparrow\downarrow}
\left(k,-k;k,-k\right)G_{\notzero\downarrow}(-k).
\label{ea18}
\end{align}

Since the restricted self-energy is zero, the restricted self-consistent
Green function equals to the bare Green function
\begin{align}
G_{\notzero\downarrow}(-k)&=G^0_\downarrow(-k).
\label{ea19}
\end{align}
Writing the only non-trivial term as 
\begin{equation}
{\mathcal{T}}_{\uparrow\downarrow}(k,-k;k,-k)=-\frac{V}{k_{\rm
B}T}\Delta^*_{\bf k}\Delta_{\bf k}
\label{ea19a}
\end{equation}
one arrives at the Gor'kov self-energy \eqref{ea13}. We have used 
$\chi(k)=0$ and $Z(k)=1$ following from \eqref{ea17} so that
$\phi(k)=\Delta_{\bf k}$.

It remains to prove that $\Delta^*_{\bf k}$ defined via 
equation \eqref{ea19} satisfies the BCS gap equation 
\eqref{ea21gap}. This directly follows from the gap 
equation \eqref{ap6} which for the potential \eqref{ea2}
reads
\begin{equation}
\Delta^*(k)=-\lambda\,\zeta_{\bf k}
\frac{k_{\rm B}T}{V}
{\sum_{k'}}\zeta_{\bf k'}
G_{\uparrow}(k')G_{{\notzero}\downarrow}(-k')~\Delta^*(k').
\label{ap6ap}
\end{equation}
The gap function does not depend on the Matsubara frequency $\omega$, with 
$k\equiv(\omega,{\bf k})$, since the interaction is not retarded so that 
$\chi$ is independent of $\omega$. Accordingly, $\Delta^*(k)=\Delta^*_{\bf k}$. 
Denoting $k'\equiv(\omega',{\bf k'})$ and using Eq.~\eqref{ea19} one 
recovers the BCS gap equation \eqref{ea21gap}. 

The restricted self-consistent T-matrix thus also yields the exact 
result for this special model. 

\bibliography{RSCtheory}

\end{document}